\documentclass[aps,prb,showpacs,twocolumn]{revtex4}

\usepackage[T1]{fontenc}
\usepackage{amssymb}
\usepackage{amsbsy}
\usepackage{amsmath}             
\usepackage{epsfig}

\begin{document}

\title{Mott insulator to superfluid transition in the Bose-Hubbard
  model: a strong-coupling approach} 

\author{K. Sengupta}
\affiliation{ 
Department of Physics, University of Toronto,
60 St. George Street, Toronto M5T 2Y4 ON, Canada \\ and 
Department of Physics, Yale university, New Haven,
CT-06520-8120  }
\author{N. Dupuis}
\affiliation{Department of Mathematics, Imperial College, 
180 Queen's Gate, London SW7 2AZ, UK \\ and 
Laboratoire de Physique des Solides, CNRS UMR 8502, \\
  Universit\'e Paris-Sud, 91405 Orsay, France }

\date{December 8, 2004}


\begin{abstract} 
We present a strong-coupling expansion of the
Bose-Hubbard model which describes both the superfluid and the Mott
phases of ultracold bosonic atoms in an optical lattice. By performing
two successive Hubbard-Stratonovich transformations of 
the intersite hopping term, we derive an effective
action which provides a suitable starting point to study the
strong-coupling limit of the Bose-Hubbard model. This action can
be analyzed by taking into account Gaussian fluctuations about
the mean-field approximation as in the Bogoliubov theory of the weakly
interacting Bose gas. In the Mott
phase, we reproduce results of previous mean-field theories and also
calculate the momentum distribution function. In the superfluid phase,
we find a gapless spectrum and compare our results with the Bogoliubov
theory.
\end{abstract}


\maketitle

\section{Introduction}

Recent experiments on ultracold trapped atomic gases have opened
a new window onto the phases of quantum matter. \cite{Greiner02,Stoferle04}
A gas of bosonic atoms in an optical or magnetic trap has been
reversibly tuned  between superfluid (SF)
and insulating ground states by varying the strength of a
periodic potential produced by standing optical waves.
This transition has been explained on the basis of the Bose-Hubbard
model with on-site repulsive interactions and hopping between
nearest neighboring sites of the lattice.
\cite{Fisher89}  As long as the atom-atom interactions are
small compared to the hopping amplitude, the ground state remains
superfluid. In the opposite limit of a strong lattice potential, the
interaction energy dominates and the ground state is a Mott insulator
(MI) when the density is commensurate, with an integer number of atoms
localized at each lattice site. 

The Gross-Pitaevskii equation or the Bogoliubov theory
\cite{Pitaevskii03} assume  quantum fluctuations to be small and are
unable to describe the SF-MI transition and the MI phase. The SF-MI
transition is usually 
studied within a strong-coupling perturbation theory which assumes the
kinetic energy to be small and treats exactly the on-site
repulsion. In the simplest version, the kinetic energy term is considered
within mean-field theory. 
\cite{Fisher89,Sheshadri93,VanOosten01,Sachdev99} The mean-field
approximation  is well known 
to give a reasonable estimate of the critical on-site
repulsion at which the MI-SF transition occurs. Fluctuation corrections to
the mean-field approach have also been considered within a systematic
strong-coupling expansion. \cite{Freericks94} All these approaches
have given a reasonable description of the MI phase and in particular
of the excitation spectrum. However, they
have not provided a description of the SF phase. 

In this work, we develop a strong-coupling expansion of the
Bose-Hubbard model which allows us to extend the treatment of
Refs.~\onlinecite{Fisher89,Sheshadri93,VanOosten01,Sachdev99} and describe
both the MI and SF phases. Our approach is similar to strong-coupling
expansions 
introduced for the (fermionic) Hubbard model. \cite{Pairault98,Dupuis00}
In Sec.~\ref{sec:Seff}, we derive an effective action for the
Bose-Hubbard model in the strong-coupling limit by performing two successive
Hubbard-Stratonovich transformations of the intersite hopping
term. This effective action involves the exact one- and two-particle
Green's functions in the local limit (i.e. in the absence of intersite
hopping). We then use the standard Bogoliubov approximation: we 
perform a saddle-point (or mean-field) approximation and expand the
action to quadratic order in the fluctuations (Sec.~\ref{sec:Bog}). In
the MI phase, we recover the previous mean-field result:  
\cite{Sheshadri93,VanOosten01} We find a gapped
excitation spectrum which becomes gapless at the MI-SF transition. We
also calculate the momentum distribution function and study 
the critical behavior at the
transition. In the SF phase, we obtain a gapless spectrum (in
agreement with Goldstone theorem) and compute the Bogoliubov sound
mode velocity. We compare our results with the Bogoliubov theory.

\section{Effective action in the strong-coupling limit}
\label{sec:Seff}

The Bose-Hubbard model is defined by the Hamiltonian
\begin{equation}
H = - t \sum_{\langle {\bf r},{\bf r}'\rangle} (\hat\psi^\dagger_{\bf r}
\hat\psi_{{\bf r}'} + {\rm h.c.}) - \mu \sum_{\bf r} \hat n_{\bf r} +
\frac{U}{2} \sum_{\bf r} \hat n_{\bf r}(\hat n_{\bf r}-1) ,
\label{ham}
\end{equation} 
where $\hat \psi_{\bf r},\hat \psi^\dagger_{\bf r}$ are bosonic operators and
$\hat n_{\bf r}=\hat\psi^\dagger_{\bf r}\hat\psi_{\bf r}$. The
discrete variable 
${\bf r}$ labels the different sites (i.e. minima) of the optical
lattice. $t$ is the hopping amplitude between nearest sites $\langle
{\bf r},{\bf r}'\rangle$ and $U$ the on-site repulsion. The optical
lattice is assumed to be bipartite with coordination number $z$. The
density, i.e. the average number $n$ of bosons per site, is fixed by
the chemical potential $\mu$.  

\begin{figure}[h]
\epsfxsize 7.cm
\epsffile[18 440 250 590]{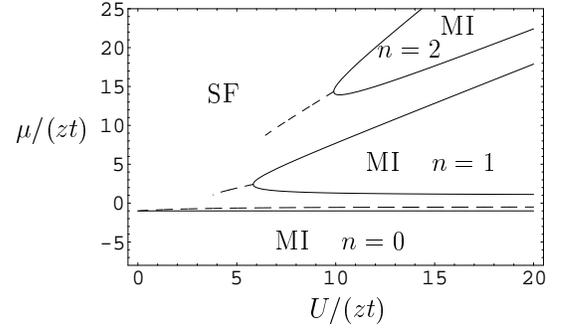}
\caption{Phase diagram of the Bose-Hubbard model showing the
  superfluid phase (SF) and the Mott insulating (MI) phases at
  commensurate filling $n$. The dashed lines corresponds to a fixed
  density $n=0.2$, $n=1$ and $n=2$. For a commensurate density $n$,
  the MI-SF transition occurs for $U/(zt)=2n+1+2(n^2+n)^{1/2}$ (for
  $n=1$, this yields $U/(zt)\simeq 5.83$, i.e. $U/t\simeq 23.31$ for a
  two-dimensional atomic gas in a square optical lattice). }
\label{fig:phase_dia} 
\end{figure}

We write the partition function $Z$ as a functional integral over a
complex field $\psi$ with the action $S[\psi^*,\psi]=\int_0^\beta d\tau \lbrace
\sum_{\bf r} \psi^*_{\bf r}\partial_\tau \psi_{\bf r} +
H[\psi^*,\psi]\rbrace$ [$\tau$ is an imaginary time and $\beta=1/T$ the
inverse temperature]. Introducing an auxialiary field $\phi$ to
decouple the intersite hopping term by means of a Hubbard-Stratonovich 
transformation, \cite{Pairault98,Dupuis00} we obtain
\begin{eqnarray}
Z &=& \int {\cal D}[\psi^*,\psi,\phi^*,\phi] e^{-(\phi|t^{-1}\phi) + 
[(\phi|\psi) + {\rm c.c.}] -S_0[\psi^*,\psi] } \nonumber \\ 
&=& Z_0 \int {\cal D}[\phi^*,\phi]  e^{-(\phi|t^{-1}\phi)} 
\bigl\langle e^{(\phi|\psi) + {\rm c.c.}} \bigr\rangle_0
\nonumber \\ &=&  Z_0 \int {\cal D}[\phi^*,\phi]
e^{-(\phi|t^{-1}\phi) +W[\phi^*,\phi] } , 
\label{Z1}
\end{eqnarray}
where we use the shorthand notation $(\phi|\psi)=\sum_a \phi^*_a\psi_a
=\int_0^\beta d\tau_a
\sum_{{\bf r}_a} \phi^*({\bf r}_a)\psi({\bf r}_a)$.
$t^{-1}$ denotes the inverse of the intersite hopping matrix
defined by $t_{{\bf r}{\bf r}'}=t$ if ${\bf r},{\bf r}'$
are nearest neighbors and $t_{{\bf r}{\bf r}'}=0$
otherwise. $S_0$ and $Z_0$ are the action and partition function in
the local limit ($t=0$). $\langle \cdots \rangle_0$ means that the
average is taken with $S_0[\psi^*,\psi]$. In the last line of
(\ref{Z1}), we have introduced the generating function 
$W[\phi^*,\phi]=\ln \langle \exp \sum_a(\phi^*_a\psi_a+{\rm c.c.}) \rangle_0$
of connected local Green's functions:\cite{Negele}
\begin{eqnarray}
G^{\rm Rc}_{\lbrace a_i,b_i \rbrace} &=& (-1)^R \langle \psi_{a_1}
  \cdots \psi_{a_R} \psi^*_{b_R}\cdots \psi^*_{b_1} \rangle \nonumber
  \\ &=& \frac{(-1)^R \delta^{(2R)}W[\phi^*,\phi]}{\delta \phi^*_{a_1}\cdots
  \delta \phi^*_{a_R} \delta \phi_{b_R}\cdots \delta \phi_{b_1}}
  \Biggr \vert_{\phi^*=\phi=0} ,
\label{GRc}
\end{eqnarray} 
where $\lbrace a_i,b_i \rbrace=\lbrace a_1\cdots a_R,b_1\cdots b_R
\rbrace$. Inverting Eq.~(\ref{GRc}), we obtain
\begin{equation}
W[\phi^*,\phi] = \sum_{R=1}^\infty \frac{(-1)^R}{(R!)^2} \sum'_{a_1\cdots
  b_R} G^{\rm Rc}_{\lbrace a_i,b_i \rbrace} \phi^*_{a_1}\cdots
  \phi^*_{a_R}\phi_{b_R}\cdots  \phi_{b_1} ,
\end{equation}
where $\sum'$ means that all the fields share the same value of the
site index. If we truncate $W[\phi^*,\phi]$ to quartic order in the fields,
we obtain the action
\begin{eqnarray}
S[\phi^*,\phi] &=& (\phi|t^{-1}\phi) - W[\phi^*,\phi] \nonumber \\ 
&=& \sum_{a,b} \phi^*_a (t^{-1}_{ab}+G_{ab}) \phi_b \nonumber \\ 
&& - \frac{1}{4} \sum_{a_1,a_2,b_1,b_2} G^{\rm IIc}_{a_1a_2,b_1b_2}
\phi^*_{a_1}\phi^*_{a_2} \phi_{b_2} \phi_{b_1} ,
\label{Sphi}
\end{eqnarray}
where $G\equiv G^{\rm I}$. 
Eq.~(\ref{Sphi}) was used as a starting point by van Oosten {\it et. al.} to
study the instability of the MI with respect to superfluidity. 
\cite{VanOosten01} Their results are summarized in Appendix
\ref{app:auxiliary} and lead to the usual mean-field phase diagram shown in 
Fig.~\ref{fig:phase_dia}. It is tempting to go
beyond the mean-field approximation by considering Gaussian
fluctuations of the $\phi$ field about its mean-field value. The
Green's function obtained in this way is however not physical
since it leads in the SF phase to a spectral function which is
not normalized to unity.\cite{note1} Physical quantities like the
excitation spectrum, the velocity of the Bogoliubov sound mode or the
momentum distribution in the SF phase are therefore out of
reach within this approach.  

These difficulties can be circumvented if one performs a second
Hubbard-Stratonovich decoupling of the hopping term: 
\begin{equation}
Z = Z_0 \int {\cal D}[\psi^*,\psi,\phi^*,\phi] e^{(\psi|t \psi) -
    [(\psi|\phi)+{\rm c.c.}] + W[\phi^*,\phi] } .
\label{Z2}
\end{equation}
In Appendix \ref{app:hs}, we show that the auxiliary field of this
transformation has the same correlation functions as the original
boson field (hence the same notation for both fields). The
effective action $S[\psi^*,\psi]$ is obtained by integrating out the $\phi$
field in Eq.~(\ref{Z2}). This procedure was carried out in detail in 
Ref.~\onlinecite{Dupuis00} in the context of the fermionic Hubbard
model. Similarly, we obtain\cite{note2} 
\begin{eqnarray}
S[\psi^*,\psi] &=& -\sum_{a,b} \psi^*_a (G^{-1}_{ab}+t_{ab}) \psi_b \nonumber
\\ && + \frac{1}{4} \sum_{a_1,a_2,b_1,b_2} \Gamma^{\rm
  II}_{a_1a_2,b_1,b_2} \psi^*_{a_1} \psi^*_{a_2} \psi_{b_2} \psi_{b_1} ,
\label{action0}
\end{eqnarray}
where $\Gamma^{\rm II}(\tau_1,\tau_2;\tau_3,\tau_4)$ is the (exact)
two-particle vertex in the local limit. In Eq.~(\ref{action0}), we
have neglected $R$-particle vertices ($R\geq 3$) whose amplitudes are
given by the (exact) local $R$-particle vertices $\Gamma^{\rm
  R}$. \cite{Dupuis00}
$\Gamma^{\rm II}$ is local in space but has a complicated time
dependence (see Appendix \ref{app:gf}). In the following, we
approximate $\Gamma^{\rm II}$ by its static value (obtained by passing
to frequency space and putting all Matsubara frequencies to
zero). This approximation is justified for energies much below $U$
where the frequency dependence of the local two-particle vertex is
weak. At higher energies, its validity is more difficult to assess. 
Introducing 
\begin{equation}
g=\frac{1}{2}\Gamma^{\rm II}|_{\rm static}, 
\end{equation}
we finally obtain
\begin{eqnarray}
S &=& - \int_0^\beta d\tau d\tau' \sum_{{\bf r},{\bf r}'} 
\psi^*_{\bf r}(\tau)[G^{-1}({\bf
  r},\tau;{\bf r}',\tau') \nonumber \\ && + t_{{\bf r},{\bf r}'}\delta
(\tau-\tau')]  \psi_{{\bf r}'}(\tau') 
+ \frac{g}{2} \int_0^\beta d\tau \sum_{\bf r} \psi^*_{\bf r}\psi^*_{\bf r} 
 \psi_{\bf r}\psi_{\bf r} . \nonumber \\ && 
\label{action1}
\end{eqnarray}
The action (\ref{action1}) is the starting point of our analysis. It
is analog to the original action  $\int_0^\beta d\tau \lbrace
\sum_{\bf r} \psi^*_{\bf r}\partial_\tau \psi_{\bf r} +
H[\psi^*,\psi]\rbrace$ with two noteworthy differences: the ``free''
propagator involves the exact local propagator $G$, and the
amplitude of the boson-boson interaction is given by the exact local
two-particle vertex (approximated here by its static limit). The
action (\ref{action1}) yields the exact partition function $Z=Z_0 \int
{\cal D}[\psi^*,\psi] e^{-S}$ and the exact Green function $-\langle
\psi_{\bf r}(\tau)\psi^*_{{\bf r}'}(\tau')\rangle$ both in the local
($t=0$) and non-interacting ($U=0$)
limits. \cite{Pairault98,Dupuis00}. By means of two successive
Hubbard-Stratonovich transformations of the intersite hopping term, 
we have thus performed a partial 
resummation of interaction processes and obtained an effective action
which provides a suitable starting point in the strong-coupling
limit.

\section{Mean-field and Gaussian approximations} 
\label{sec:Bog}

In order to study the Mott and superfluid phases from the strong-coupling
effective action (\ref{action1}), we use the standard Bogoliubov
approximation: we first perform a saddle-point (or mean-field)
approximation and then expand the action (\ref{action1}) to 
quadratic order in the fluctuations. The saddle-point action is given by
\begin{equation}
\frac{S}{N\beta} = -(\bar G^{-1}+D)\psi_0^2+\frac{g}{2} \psi_0^4,
\end{equation}
where $\bar G=G(i\omega=0)$, $D=zt$, and $N$ is the total number of
lattice sites. The saddle-point value $\psi_0$ (assumed here, with no
loss of generality, to be real) is obtained from $\partial S/\partial
\psi_0=0$: 
\begin{equation}
\psi_0^2=
\left\lbrace 
\begin{array}{l}
\dfrac{\bar G^{-1}+D}{g} \,\,\, {\rm if}\,\,\, \bar G^{-1}+D>0, \\
0 \,\,\, {\rm otherwise.}
\end{array}
\right .  
\end{equation}
The MI-SF therefore occurs when $\bar G^{-1}+D=0$, in
agreement with the results of Appendix \ref{app:auxiliary}, which leads to
the phase diagram shown in Fig.~\ref{fig:phase_dia}. Using $\langle
\psi_{\bf r}\rangle = \delta \ln Z(J^*,J)/\delta J^*_{\bf
  r}|_{J^*=J=0}$, where $Z[J^*,J]$ is given by Eq.~(\ref{ZJ1}) of appendix
\ref{app:hs}, we obtain $\phi_0=D\psi_0$ where
$\phi_0$ is the mean value of the auxiliary field. Near the MI-SF
transition, where $\bar G^{-1}+D \approx 0$, we then find $\phi_0^2
\simeq 2(D^{-1}+\bar G)/\bar G^{\rm IIc}$ in agreement with the result
of Appendix \ref{app:auxiliary}. 

To quadratic order in the fluctuations $\tilde\psi_{\bf r}=\psi_{\bf
  r}-\psi_0$, we obtain the action
\begin{widetext}
\begin{equation}
S=\frac{1}{2} \sum_{{\bf k},\omega} (\tilde\psi^*({\bf k},i\omega),
\tilde\psi(-{\bf k},-i\omega))  
\left(
\begin{array}{lr}
-G^{-1}(i\omega)+\epsilon_{\bf k}+2g\psi_0^2 & g \psi_0^2 \\
g \psi_0^2 & -G^{-1}(-i\omega)+\epsilon_{-{\bf k}}+2g\psi_0^2
\end{array}
\right)
\left(
\begin{array}{l}
\tilde \psi({\bf k},i\omega) \\ 
\tilde \psi^*(-{\bf k},-i\omega)
\end{array}
\right) ,
\label{action2}
\end{equation}
\end{widetext}
where $\tilde \psi({\bf k},i\omega)$ is the Fourier transformed field
of $\tilde\psi_{\bf r}(\tau)$ and $\omega$ a bosonic Matsubara
frequency. $\epsilon_{\bf k}$, the Fourier transform of $-t_{{\bf
    r},{\bf r}'}$, is the boson dispersion in the absence of
the one-site repulsion.

\subsection{Mott phase and the MI-SF transition} 
\label{subsec:MI}

In the Mott phase, where $\psi_0=0$, the Green's function ${\cal
  G}({\bf k},i\omega)=-\langle \psi({\bf
  k},i\omega) \psi^*({\bf k},i\omega)\rangle$ can be directly read
  off from Eq.~(\ref{action2}): ${\cal G}^{-1}({\bf
  k},i\omega)=G^{-1}(i\omega) - \epsilon_{\bf k}$. 
Using Eq.~(\ref{G1}), one obtains
\begin{equation}
{\cal G}({\bf k},i\omega)=
\frac{1-z_{\bf k}}{i\omega-E^-_{\bf k}} + \frac{z_{\bf k}}{i\omega-E^+_{\bf
    k}} .
\label{gf1}
\end{equation}
The two excitation energies $E^\pm_{\bf k}$
and the spectral weight $z_{\bf k}$ 
are defined by
\begin{eqnarray}
E^\pm_{\bf k} &=& -\delta \mu + \frac{\epsilon_{\bf k}}{2} \pm \frac{1}{2}
\Bigl[\epsilon_{\bf k}^2+4\epsilon_{\bf k}Ux+U^2 \Bigr]^{1/2} , \nonumber \\
z_{\bf k} &=& \frac{E^+_{\bf k}+\delta\mu+Ux}{E^+_{\bf k}-E^-_{\bf k}} ,
\end{eqnarray}
where $x=n_0+1/2$ and $\delta\mu=\mu-U(n_0-1/2)$. $n_0\equiv n_0(\mu)$ is the
(integer) number of bosons in the local limit for a chemical potential
$\mu$ (see Appendix \ref{app:gf}).  

\begin{figure}[h]
\epsfxsize 7cm
\epsffile[40 445 280 585]{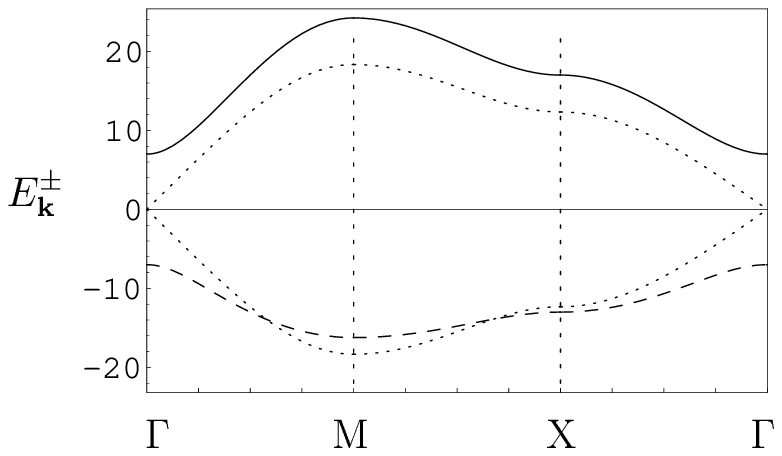}
\epsfxsize 7cm
\epsffile[40 445 280 585]{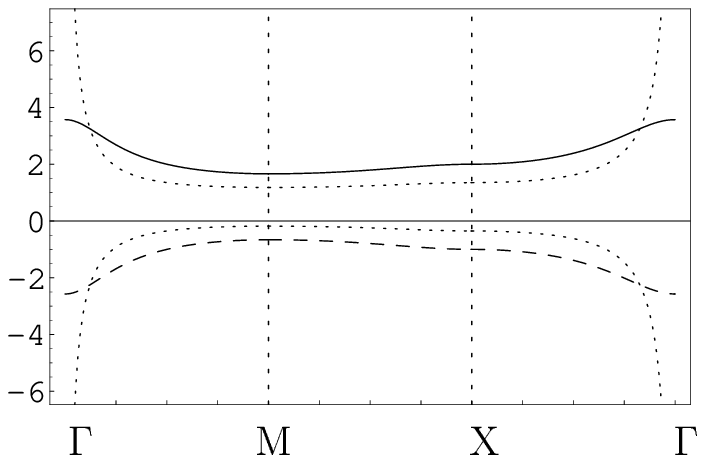}
\caption{Top: Excitation energies $E^+_{\bf k}$ (solid line) and
  $E^-_{\bf k}$ (dashed line) in the  MI $n=1$ for
  $U=30t$. Bottom: Spectral weight $z_{\bf k}$ (solid line) and
  $1-z_{\bf k}$ (dashed line). The dotted lines show the result obtained
  from the Bogoliubov theory (which predicts the phase to be
  superfluid).  [$\Gamma=(0,0)$, ${\rm M}=(\pi,\pi)$ 
  and ${\rm X}=(\pi,0)$.] Results shown in
  Figs.~\ref{fig:Emott1}-\ref{fig:Esf1} are obtained for a two-dimensional
  atomic gas in a square optical lattice. }
\label{fig:Emott1}
\epsfxsize 7cm
\epsffile[40 445 280 585]{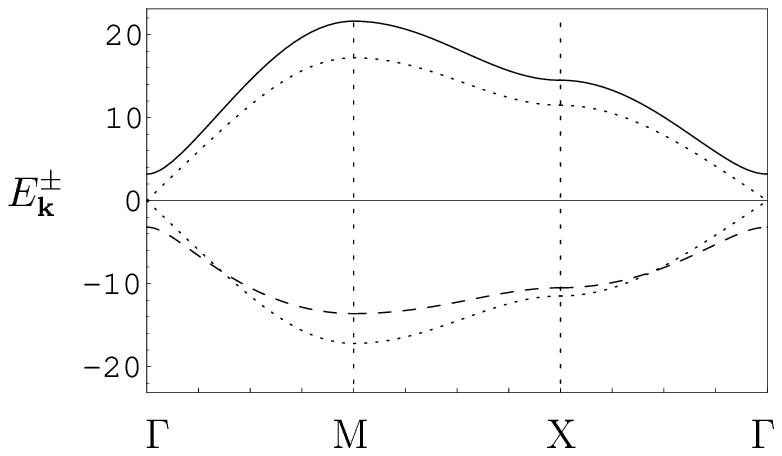}
\epsfxsize 7cm
\epsffile[40 445 280 585]{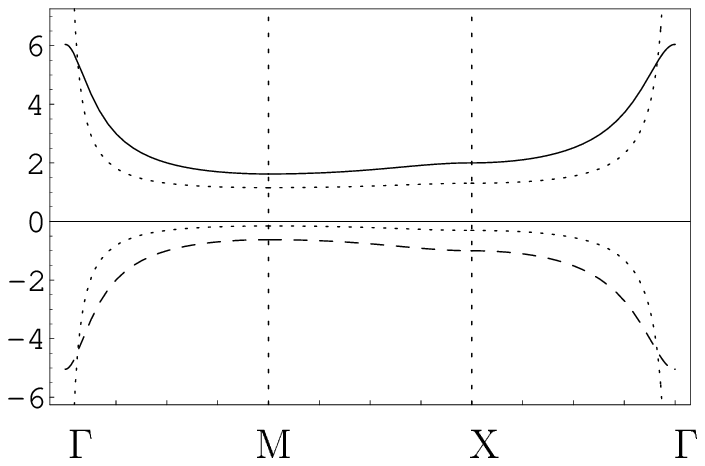}
\caption{Same as Fig.~\ref{fig:Emott1}, but for $U=25t$. }
\label{fig:Emott2}
\end{figure}

The excitation energies $E^+_{\bf k}$,$E^-_{\bf k}$, and the
corresponding spectral weight $z_{\bf k}$ and $1-z_{\bf k}$, are shown
in Figs.~\ref{fig:Emott1}-\ref{fig:Emott2} in the MI $n=1$ of a
two-dimensional atomic gas in a square optical lattice. The
spectrum exhibits a gap $E^+_{{\bf k}=0}-E^-_{{\bf
k}=0}=(D^2-4DUx+U^2)^{1/2}$ which decreases as $U$ decreases. The MI
becomes unstable against superfluidity when $E^+_{{\bf k}=0}=0$ or
$E^-_{{\bf k}=0}=0$, which agrees with Eq.~(\ref{mu_mf}) of Appendix
\ref{app:auxiliary} and leads to the phase diagram shown in
Fig.~\ref{fig:phase_dia}. The gap $E^+_{{\bf k}=0}-E^-_{{\bf
    k}=0}=(D^2-4DUx+U^2)^{1/2}$ closes at the
transition if both $E^+_{{\bf k}=0}$ and $E^-_{{\bf k}=0}$ vanish, which
occurs at the tip of the Mott lob. The MI-SF transition then takes
place at fixed density, which is the situation of physical
interest. Figs.~\ref{fig:Emott1}-\ref{fig:Emott2} are 
obtained with a chemical potential $\delta\mu=-D/2$, which ensures
that the MI-SF transition takes place at fixed density $n=1$ (see
Appendix \ref{app:auxiliary}).  
The decreasing of the Mott gap is accompanied by an increase of
spectral weight at ${\bf k}=0$, which diverges at the transition. 
Figs.~\ref{fig:Emott1}-\ref{fig:Emott2} also 
show the results of the Bogoliubov theory (as applied to the original
Hamiltonian (\ref{ham})). The Bogoliubov theory always predicts the 
ground-state to be superfluid. \cite{VanOosten01} Away from ${\bf k}=0$, it
provides a good approximation of the negative energy branch $E^-_{\bf
k}$ but gives a poor description of $E^+_{\bf k}$.

If we expand the equation $E^\pm_{{\bf k}=0}=0$ to order $O(t^2/U)$, we
obtain  
\begin{eqnarray}
\mu-Un_0+D(n_0+1)+\frac{D^2}{U}(n_0^2+n_0) &=& 0 , \nonumber \\ 
\mu-U(n_0-1)-Dn_0-\frac{D^2}{U}(n_0^2+n_0) &=& 0 ,
\end{eqnarray}
which differs from the energy calculation of
Ref.~\onlinecite{Freericks94} by terms of order $O(t^2/U)$. This
discrepancy results from the 
neglect of the one-loop correction due to $\Gamma^{\rm II}$ in the
calculation of the Green's function [Eq.~(\ref{gf1})], which also gives a
contribution of order $O(t^2/U)$. However, even without this term the
phase diagram looks qualitatively similar to the Freericks and Monien
phase diagram. 

\begin{figure}
\epsfxsize 7cm
\epsffile[40 435 280 605]{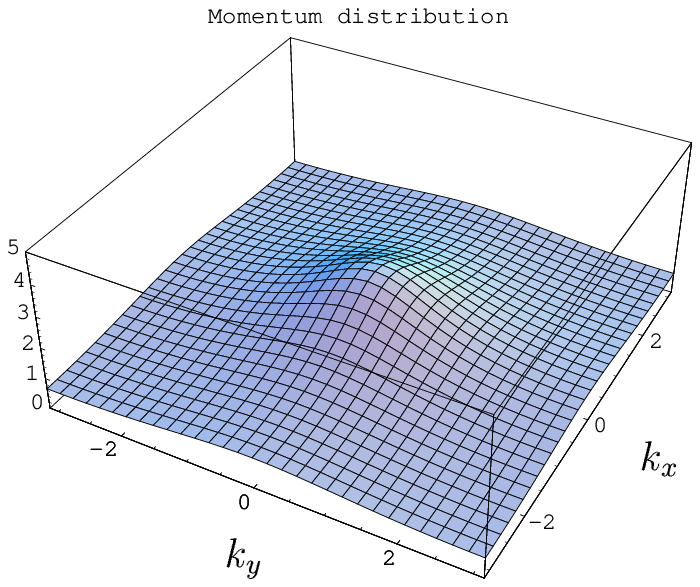}
\epsfxsize 7cm
\epsffile[40 435 280 605]{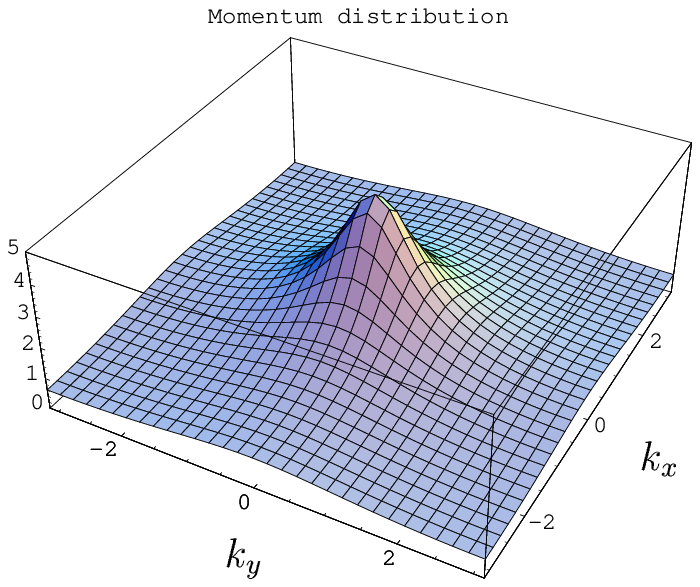}
\caption{Momentum distribution $n_{\bf k}=\langle \psi^*_{\bf
    k}\psi_{\bf k}\rangle$ in the MI $n=1$ for $U=30t$
    (top) and $U=25$ (bottom). }
\label{fig:md}
\end{figure}

From the Green's function (\ref{gf1}), we can also obtain the momentum 
distribution $n_{\bf k}=\langle \psi^*_{\bf k}\psi_{\bf k}\rangle =
-\int_{-\infty}^0 d\omega A({\bf k},\omega)=1-z_{\bf k}$. $n_{\bf k}$
measures the spectral weight of the negative energy $E^-_{\bf k}$ of
the spectrum. Deep in the Mott phase, the momentum distribution is
roughly flat. Closer to the MI-SF transition, a
peak develops around ${\bf k}=0$. This peak diverges at the
transition (Fig.~\ref{fig:md}). 

The critical theory of the SF-MI transition can be obtained from the
action (\ref{action1}) by expanding the inverse propagator
$G^{-1}(i\omega)-\epsilon_{\bf k}$ to quadratic order in ${\bf k}$ and
$\omega$. Noting that $\partial G^{-1}(i\omega)/\partial
(i\omega)|_{i\omega =0}=\partial \bar G^{-1}/\partial \mu$ (and
similarly for the second-order derivative), we obtain
\begin{eqnarray}
S &=& \int_0^\beta d\tau \int dr \Bigl [r_0 |\psi_{\bf r}^2| +K_1
\psi^*_{\bf r}\partial_\tau \psi_{\bf r} + K_2 |\partial_\tau
\psi_{\bf r}|^2 \nonumber \\ 
&& + K_3
|\boldsymbol{\nabla} \psi_{\bf r}|^2 +\frac{u}{2} |\psi_{\bf r}|^4 \Bigr],
\end{eqnarray}
where
\begin{eqnarray}
r_0 &\propto & \bar G^{-1}+D , \nonumber \\
K_1 & \propto & \frac{\partial r_0}{\partial \mu} .
\end{eqnarray} 
At all points on the MI-SF transition line except at the Mott lob tip,
$r_0$ vanishes but $K_1$
remains finite. The critical theory has then a dynamical exponent
$z=2$. At the tip of the Mott lob where both $r_0$ and $K_1$ vanish, 
the dynamical exponent $z=1$. A similar
analysis, based on the effective action $S[\phi^*,\phi]$, can be found
in Ref.~\onlinecite{Sachdev99}. 

\subsection{Superfluid phase}

\begin{figure}[h]
\epsfxsize 7cm
\epsffile[40 445 280 585]{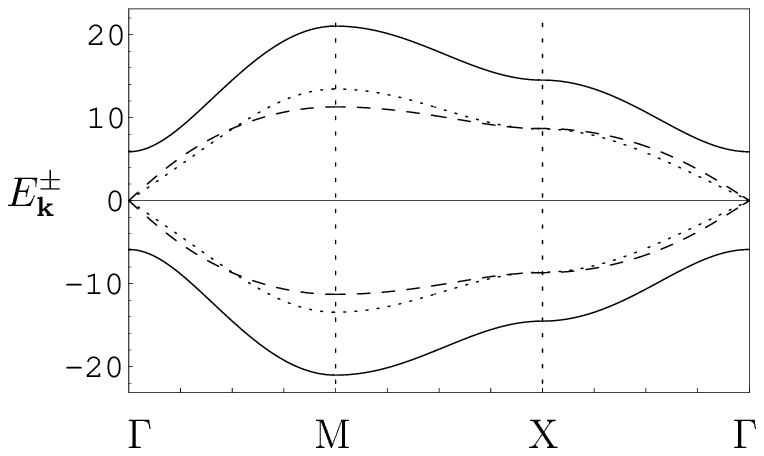}
\epsfxsize 7cm
\epsffile[40 445 280 585]{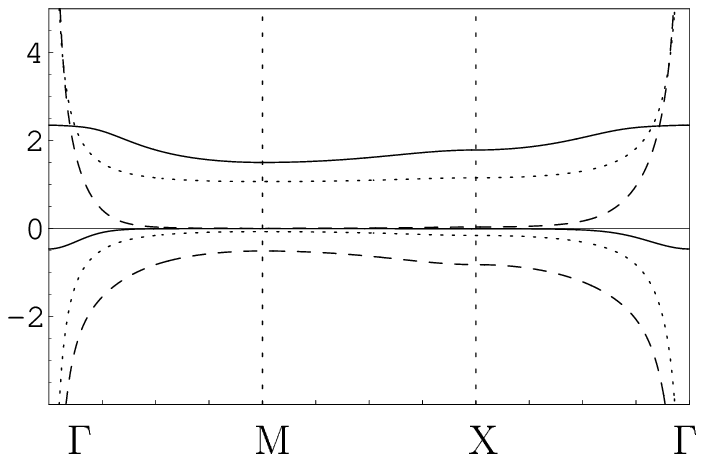}
\caption{Excitation energies $\pm E^\pm_{\bf k}$ and 
  spectral weight in the  SF phase $n=1$ and
  $U=20$. } 
\label{fig:Esf1}
\end{figure}

In the SF phase ($\psi_0\neq 0$), 
the Green's function of the $\tilde\psi$ field is obtained by inverting the
$2\times 2$ matrix propagator in Eq.~(\ref{action2}). For the diagonal
component ${\cal
  G}({\bf k},i\omega)=-\langle \tilde\psi({\bf
  k},i\omega)\tilde\psi^*({\bf k},i\omega)\rangle$, we obtain
\begin{equation}
{\cal G}({\bf k},i\omega) = \frac{(i\omega+\delta\mu+Ux)
  (i\omega-z^+_{\bf k})(i\omega-z^-_{\bf k})}{(\omega^2+E^{+2}_{\bf
  k}) (\omega^2+E^{-2}_{\bf k})},
\label{gf2}
\end{equation}
where
\begin{eqnarray}
E^{\pm 2}_{\bf k} &=& - \frac{B_{\bf k}}{2} \pm \frac{1}{2} (B^2_{\bf
  k}-4C_{\bf k})^{1/2} , \nonumber \\
z^\pm_{\bf k} &=& \frac{\tilde A_{\bf k}}{2} \pm \frac{1}{2} (\tilde
  A_{\bf k}^2-4 \tilde B_{\bf k})^{1/2}, \nonumber \\
\tilde A_{\bf k} &=& 2\delta \mu - 2(\bar G^{-1}+D)-\epsilon_{\bf k}, \nonumber
  \\
\tilde B_{\bf k} &=& -(2\bar G^{-1}+2D+\epsilon_{\bf
  k})(\delta\mu+Ux)+\delta\mu^2 - \frac{U^2}{4}, \nonumber \\
B_{\bf k} &=& 2\tilde B_{\bf k}-\tilde A^2_{\bf k}+(\bar G^{-1}+D)^2, \nonumber
  \\
C_{\bf k} &=& \tilde B^2_{\bf k}-(\bar G^{-1}+D)^2(\delta\mu+Ux)^2 .
\end{eqnarray}
From (\ref{gf2}), we deduce the spectral function $A({\bf
k},\omega)=-\frac{1}{\pi}{\rm Im}{\cal G}({\bf k},\omega+i0^+)$: 
\begin{eqnarray}
&& A({\bf k},\omega) = \nonumber \\  
&& \frac{(E^+_{\bf k}+\delta \mu+Ux)  
(E^+_{\bf k}-z^+_{\bf k})(E^+_{\bf k}-z^-_{\bf k})}
{2E^+_{\bf k}(E^{+2}_{\bf k}-E^{-2}_{\bf k})} \delta(\omega-E^+_{\bf k})
\nonumber \\ &&
+ \frac{(E^+_{\bf k}-\delta \mu-Ux)  
(E^+_{\bf k}+z^+_{\bf k})(E^+_{\bf k}+z^-_{\bf k})}
{2E^+_{\bf k}(E^{+2}_{\bf k}-E^{-2}_{\bf k})} \delta(\omega+E^+_{\bf
  k})
\nonumber \\ &&
-\frac{(E^-_{\bf k}+\delta \mu+Ux)  
(E^-_{\bf k}-z^+_{\bf k})(E^-_{\bf k}-z^-_{\bf k})}
{2E^-_{\bf k}(E^{+2}_{\bf k}-E^{-2}_{\bf k})} \delta(\omega-E^-_{\bf k})
\nonumber \\ && 
-\frac{(E^-_{\bf k}-\delta \mu-Ux)  
(E^-_{\bf k}+z^+_{\bf k})(E^-_{\bf k}+z^-_{\bf k})}
{2E^-_{\bf k}(E^{+2}_{\bf k}-E^{-2}_{\bf k})} \delta(\omega+E^-_{\bf
  k}) . \nonumber \\ && 
\end{eqnarray}
The Green's function (\ref{gf2}) has the desired physical
properties. The spectral function is normalized,
$\int_{-\infty}^\infty d\omega A({\bf k},\omega)=1$, and has the
correct sign: ${\rm sgn}[A({\bf k},\omega)]={\rm
  sgn}(\omega)$. \cite{note1} There
are four excitation branches $\pm E^\pm_{\bf k}$, two of which ($\pm
E^-_{\bf k}$) being gapless for ${\bf k}\to 0$
(Fig.~\ref{fig:Esf1}). However, for a given value of ${\bf k}$, only
two branches carry a significant spectral weight. Away from
${\bf k}=0$, the spectral weight is almost completely exhausted by
$E^+_{\bf k}$ and $-E^-_{\bf k}$. In the vicinity of ${\bf k}=0$, the
two gapless branches $\pm E^-_{\bf k}$ exhaust the spectral weight. 
By expanding $E^-_{\bf k}$ in the vicinity of ${\bf k}\to 0$, we find
a linear spectrum 
\begin{equation}
E^-_{\bf k} = c|{\bf k}|,
\end{equation}
where
\begin{eqnarray}
c &=& \left[\frac{2t(\bar G^{-1}+D)}{\alpha^2+2\gamma(\bar G^{-1}+D)}
\right]^{1/2} ,
\nonumber \\
\alpha &=& \frac{\delta\mu^2+2\delta\mu Ux+U^2/4}{(\delta\mu+Ux)^2} ,
\nonumber \\ 
\gamma &=& \frac{U^2(x^2-1/4)}{(\delta\mu+Ux)^3} . 
\label{velocity}
\end{eqnarray}
Our strong-coupling approach therefore reproduces the Bogoliubov
(Goldstone) mode of the SF phase.

As discussed in Sec.~\ref{subsec:MI}, our strong-coupling theory is
not an expansion order by order in $t/U$. For this reason, the
computation of the chemical potential from the single-particle Green's
function, i.e. $n={\rm Tr}({\cal G})$, is not reliable. We have therefore
used the chemical potential obtained within the mean-field
approximation discussed in Appendix \ref{app:auxiliary}. 

Fig.~\ref{fig:Esf1} also shows the results of the
Bogoliubov theory (as applied to the Hamiltonian (\ref{ham})) for the
{\it same} chemical potential $\mu$. The
Bogoliubov theory provides a good approximation to $E^-_{\bf k}$ and
therefore to the low-energy part of the excitation spectrum. This
implies that the velocity of the gapless mode  [Eq.~(\ref{velocity})]
can be approximated by the Bogoliubov result $c=[2t(\mu+D)]^{1/2}$. 
Away from ${\bf k}=0$, the Bogoliubov approach gives a rather poor
description of $E^+_{\bf k}$. 

The Green's function ${\cal G}({\bf k},i\omega)$ yields the momentum
distribution 
\begin{eqnarray} 
n_{\bf k} &=& \langle \psi^*_{\bf k}\psi_{\bf k}\rangle \nonumber \\
&=& N\psi^2_0 \delta_{{\bf k},0}-\int_{-\infty}^0 d\omega A({\bf
  k},\omega), 
\end{eqnarray}
Apart from the
condensate contribution $ N\psi^2_0 \delta_{{\bf k},0}$, the momentum
distribution function is directly given by the spectral weight of the
negatives energies $-E^+_{\bf k}$ and $-E^-_{\bf k}$
(Fig.~\ref{fig:Esf1}). 

\begin{figure}[t]
\epsfxsize 7.5cm
\epsffile[20 20 270 225]{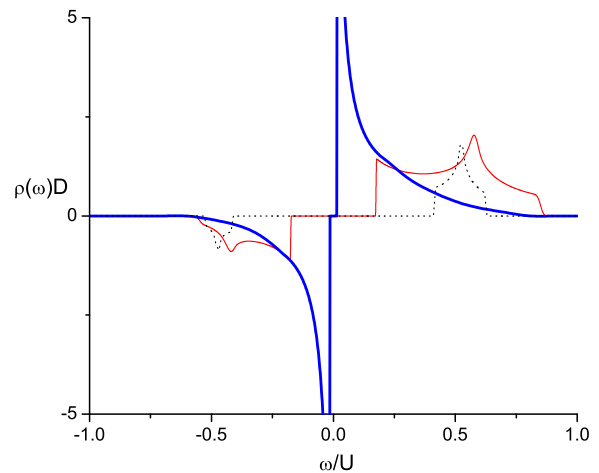}
\caption{Integrated spectral function $\rho(\omega)= \int
  \frac{d^2k}{(2\pi)^2} A({\bf k},\omega)$ in the MI $n=1$ ($\mu=U/2-D/2$):
  $U/t=80$ 
  (dashed line), 40 (thin solid line)and 23.33 (thick solid line). The
  transition to the SF phase 
  occurs for $U/t\simeq 23.31$.    
}
\label{fig:dos} 
\end{figure}

Fig.~\ref{fig:dos} shows the integrated spectral function 
$\rho(\omega)= \int \frac{d^2k}{(2\pi)^2} A({\bf k},\omega)$ for a
commensurate density $n=1$. 
Deep in the Mott phase, $\rho(\omega)$ is essentially given by the
non-interaction density of states of free bosons on the square lattice
centered around $-\mu$ and $U-\mu$ and with relative spectral weigths
$-n_0$ and $n_0+1$. The two peaks near $\omega=-\mu$ and $\omega=U-\mu$
are due to the Van Hove singularities in the density of states of free
bosons. When decreasing the value of $U/t$, the Mott gap decreases
and $\rho(\omega)$ strongly increases at the gap edges. At the critical value
$U/t\simeq 23.31$, the gap closes and $\rho(\omega)$ diverges at
$\omega=0$. This divergence persists in the superfluid phase.

\section{Conclusion}

By performing two successive Hubbard-Stratonovich transformations of
the intersite hopping term, we have shown how to derive an effective
action which provides a suitable starting point to study the
strong-coupling limit of the Bose-Hubbard model. This action can
then be analyzed by taking into account Gaussian fluctuations about
the mean-field approximation as in the Bogoliubov theory of the weakly
interacting Bose gas. The main improvement over previous related approaches
\cite{Sheshadri93,VanOosten01,Sachdev99,Freericks94} is the
possibility to describe both the Mott and SF phases. Both in the Mott
and SF phases, we compute the excitation spectrum and the momentum
distribution. Our approach clearly shows how the excitation spectrum,
which is gapped in the MI phase, becomes gapless at the MI-SF
transition. 

The strong-coupling expansion presented in this paper should in
principle also applies to more complicated situations where for
instance several atom species are present in the optical lattice. 

\vspace{0.25cm}

\noindent
{\it Note added:} after completing this paper, we became aware of two related
  works. Konabe {\it et al.} \cite{Konabe04} have studied the single-particle
  excitation spectrum in the Mott phase and obtained results similar to
  ours. The method used by these authors bears some similarities with the
  strong-coupling expansion discussed in the present paper. Within a
  slave-boson representation of the Bose-Hubbard model, Dickerscheid {\it
  et. al.} \cite{Dickerscheid03} have discussed both the Mott and SF
  phases. Their results agree with ours (whenever the comparison is possible).

\appendix

\section{Hubbard-Stratonovich transformations}
\label{app:hs} 

The Green's functions of the boson field $\psi$ can be obtained from
the generating function \cite{Negele} 
\begin{equation}
Z[J^*,J] = \int {\cal D}[\psi^*,\psi]
    e^{(\psi|t\psi)-S_0[\psi^*,\psi]+[(J|\psi)+{\rm c.c.}]}, 
\label{ZJ1}
\end{equation} 
where $J^*_{\bf r},J_{\bf r}$ are external sources. After the
Hubbard-Stratonovich decoupling of the intersite hopping term [see
Eq.~(\ref{Z1})] and the shift $\phi^*\to\phi^*-J^*$,$\phi\to\phi-J$ of
the auxiliary field, we obtain
\begin{widetext}
\begin{eqnarray}
Z[J^*,J] &=&  \int {\cal D}[\psi^*,\psi,\phi^*,\phi]
e^{-(\phi-J|t^{-1}(\phi-J))+[(\phi|\psi)+{\rm c.c.}] - S_0[\psi^*,\psi]}
  \nonumber \\ 
&=&  Z_0 \int {\cal D}[\phi^*,\phi] e^{-(\phi-J|t^{-1}(\phi-J)) +
  W[\phi^*,\phi] } . 
\end{eqnarray}
A second Hubbard-Stratonovich decoupling of the hopping term (with an
auxiliary field $\psi'$) leads to
\begin{eqnarray}
Z[J^*,J] &=& Z_0 \int {\cal D}[{\psi'}^*\psi',\phi^*,\phi] e^{
  (\psi'|t\psi')-[(\psi'|\phi-J)+{\rm c.c.}] + W[\phi^*,\phi]} \nonumber \\
&=&  Z_0 \int {\cal D}[{\psi'}^*\psi',\phi^*,\phi] e^{
  (\psi'|t\psi')-[(\psi'|\phi)+{\rm c.c.}] + [(\psi'|J)+{\rm c.c.}]
  +W[\phi^*,\phi]}. 
\label{ZJ2}
\end{eqnarray}
\end{widetext}
From (\ref{ZJ2}) we deduce that $Z[J^*,J]$ is also the generating function
of the Green's functions of the $\psi'$ field. $\psi'$ can therefore
be identified with the original boson field $\psi$.

\section{Calculation of the local Green's functions $G$ and $G^{\rm II}$}
\label{app:gf}

In the absence of intersite hopping ($t=0$), the states
$|p\rangle=(p!)^{-1/2}(\hat\psi^\dagger)^p|0\rangle$ ($p\geq 0$
integer) are eigenstates with
eigenvalues $\epsilon_p=-\mu p+(U/2)p(p-1)$. [We consider a single
site and therefore drop the site index.] $|0\rangle$ is the vacuum
of particles. This yields the partition function $Z_0=\sum_{p=0}^\infty
e^{-\beta\epsilon_p}$. In the ground-state, for a given value of the
chemical potential $\mu$, there are $n_0$
bosons per site, where $n_0$ is obtained from $\epsilon_{n_0}
={\rm min}_p \epsilon_p$. The latter condition leads to $n_0-1\leq
\mu/U \leq n_0$ if $\mu \geq -U$, and $n_0=0$ if $\mu \leq -U$. Note
that $n_0$ is integer (except when $\mu/U=p$ is integer; the states 
$|p\rangle$ and $|p+1\rangle$ are then degenerate), even when the
boson density $n$ is not.  

The single-particle Green's function $G(\tau)=-\langle T_\tau
\hat\psi(\tau) \hat\psi^\dagger(0)\rangle$ is easily calculated using
the closure relation $\sum_{p=0}^\infty |p\rangle\langle p|=1$. For
$\tau>0$, one finds
\begin{equation}
G(\tau) = -\frac{1}{Z_0}\sum_{p=0}^\infty (p+1)
e^{-(\beta-\tau)\epsilon_p-\tau \epsilon_{p+1}} ,
\end{equation}
and, in frequency space,
\begin{equation}
G(i\omega) = \frac{-n_0}{i\omega+\epsilon_{n_0-1}-
  \epsilon_{n_0}} 
+ \frac{n_0+1}{i\omega+\epsilon_{n_0}-\epsilon_{n_0+1}} ,
\label{G1}
\end{equation}
where $\omega$ is a bosonic Matsubara frequency.

The two-particle Green's function can be calculated in the same
way. One finds
\begin{widetext}
\begin{eqnarray}
G^{\rm II}(\tau_1,\tau_2;\tau_3,\tau_4=0) &=& \langle T_\tau
\hat\psi(\tau_1)\hat\psi(\tau_2) 
\hat\psi^\dagger(0)\hat\psi^\dagger(\tau_3) \rangle \nonumber \\
&=& \frac{1}{Z_0}
\sum_{p=0}^\infty e^{-\beta \epsilon_p}
\Bigl[ (p+1)(p+2) e^{\tau_1(\epsilon_p-\epsilon_{p+1})
                    +\tau_2(\epsilon_{p+1}-\epsilon_{p+2})
                    +\tau_3(\epsilon_{p+2}-\epsilon_{p+1})}
                    \theta(\tau_1-\tau_2)\theta(\tau_2-\tau_3)
                    \nonumber \\ && 
     + (p+1)(p+2) e^{\tau_1(\epsilon_{p+1}-\epsilon_{p+2})
                    +\tau_2(\epsilon_{p}-\epsilon_{p+1})
                    +\tau_3(\epsilon_{p+2}-\epsilon_{p+1})}
                    \theta(\tau_2-\tau_1)\theta(\tau_1-\tau_3)
                    \nonumber \\ && 
     + (p+1)^2    e^{\tau_1(\epsilon_{p}-\epsilon_{p+1})
                    +\tau_2(\epsilon_{p}-\epsilon_{p+1})
                    +\tau_3(\epsilon_{p+1}-\epsilon_{p})}
                    [\theta(\tau_1-\tau_3)\theta(\tau_3-\tau_2)
                     + \theta(\tau_2-\tau_3)\theta(\tau_3-\tau_1)]
                    \nonumber \\ && 
     + p(p+1)     e^{\tau_1(\epsilon_{p-1}-\epsilon_{p})
                    +\tau_2(\epsilon_{p}-\epsilon_{p+1})
                    +\tau_3(\epsilon_{p}-\epsilon_{p-1})}
                    \theta(\tau_3-\tau_1)\theta(\tau_1-\tau_2)
                    \nonumber \\ && 
     + p(p+1)     e^{\tau_1(\epsilon_{p}-\epsilon_{p+1})
                    +\tau_2(\epsilon_{p-1}-\epsilon_{p})
                    +\tau_3(\epsilon_{p}-\epsilon_{p-1})}
                    \theta(\tau_3-\tau_2)\theta(\tau_2-\tau_1) \Bigr] .
\end{eqnarray}
\end{widetext}
After a somewhat tedious calculation, we obtain for the Fourier transform
of the connected part in the static limit:
\begin{eqnarray}
\bar G^{\rm IIc} &=& \int_0^\beta d\tau_1d\tau_2d\tau_3 G^{\rm
  II}(\tau_1,\tau_2;\tau_3,0) -2\beta [G(i\omega=0)]^2 \nonumber \\ 
&=& - \frac{4(n_0+1)(n_0+2)}{(2\mu-(2n_0+1)U)(Un_0-\mu)^2} \nonumber \\ &&
-  \frac{4n_0(n_0-1)}{(\mu-U(n_0
  -1))^2(U(2n_0-3)-2\mu)} \nonumber \\ &&
+ \frac{4n_0(n_0+1)}{(\mu-Un_0)(-\mu+U(n_0-1))^2} \nonumber \\ && 
+ \frac{4n_0(n_0+1)}{(\mu-Un_0)^2(-\mu+U(n_0-1))} \nonumber \\ &&
+ \frac{4n^2_0}{(-\mu+U(n_0-1))^3} \nonumber \\ &&
+ \frac{4(n_0+1)^2}{(\mu-Un_0)^3} .
\end{eqnarray}
The static limit of the two-particle vertex $\Gamma^{\rm II}$ is equal
to $-\bar G^{\rm IIc}/\bar G^4$.

\section{Auxiliary-field mean-field approach} 
\label{app:auxiliary}

In this appendix, we review the mean-field results obtained from the
action $S[\phi^*,\phi]$ [Eq.~(\ref{Sphi})]. \cite{VanOosten01} Within a
saddle-point approximation, where the field $\phi_0$ is taken real and
assumed to be time and space independent, we action becomes
\begin{equation}
\frac{S}{N\beta} = (D^{-1}+\bar G)\phi_0^2 - \frac{1}{4} \bar
  G^{\rm IIc}\phi_0^4, 
\end{equation}
where $D=zt$. $\bar G$ and $\bar G^{\rm IIc}$ are the single-particle and
two-particle local Green's functions in the static limit (see Appendix
\ref{app:gf}). The ground-state energy per site $E=-\lim_{\beta\to\infty}
\frac{1}{N\beta}\ln Z$ is then given by [see Eq.~(\ref{Z1})]  
\begin{equation}
E = a_0+a_2\phi_0^2+a_4\phi_0^4,
\end{equation}
where $a_0=-\lim_{\beta\to\infty} \frac{1}{N\beta}\ln Z_0$ is the
ground-state energy in the 
local limit, $a_2=D^{-1}+\bar G$, and $a_4=-\frac{1}{4}\bar G^{\rm
IIc}$. The mean-field value $\phi_0$ is
obtained by minimizing $E$. $\phi_0$ vanishes in the Mott phase
($a_2>0$) and takes a finite value in the SF phase
($a_2<0$). The MI-SF transition is then given by
$a_2=0$, which leads to 
\begin{equation}
\delta \mu_\pm = -\frac{D}{2} \pm \frac{1}{2}
\Bigl[D^2+U^2-4DUx \Bigr]^{1/2} ,
\label{mu_mf} 
\end{equation}
where $n_0$ is the integer number of bosons in the local limit for a
chemical potential $\mu$ (see Appendix \ref{app:gf}). $x$ and
$\delta\mu$ are defined in Sec.~\ref{sec:Bog}. For each value of
$n_0$, Eq.~(\ref{mu_mf}) defines a Mott lob in the $U-\mu$ phase
diagram (Fig.~\ref{fig:phase_dia}), whose tip corresponds to
$\delta\mu_+=\delta\mu_-=-zt/2$ and
$U/(zt)=2n_0+1+2(n_0^2+n_0)^{1/2}$. At the lob tip, $\partial
a_2/\partial \mu=0$. 

In the SF phase, the order parameter $\phi_0$ is
given by $\phi_0^2=-a_2/(2a_4)$, and the ground-state energy takes the
value 
\begin{equation}
E=a_0-\frac{a_2^2}{4a_4} .
\label{Emf}
\end{equation}
From (\ref{Emf}), we deduce the mean boson density 
\begin{equation}
n = - \frac{\partial E}{\partial \mu} = n_0 + \frac{1}{4}
\frac{\partial}{\partial\mu} \left(\frac{a_2^2}{a_4}\right) 
\simeq  n_0 + \frac{a_2}{2a_4}
\frac{\partial a_2}{\partial\mu} ,
\end{equation}
where the last equality holds near the MI-SF
transition ($a_2 \approx 0$). We have used $n_0 = - \partial
a_0/\partial \mu$. We conclude that, at the MI-SF transition, the
boson density remains pinned at the integer value $n_0$ if
$\partial a_2/\partial\mu=0$, which corresponds to the tip of the Mott
lob in the $\mu-U$ phase diagram (Fig.~\ref{fig:phase_dia}).


\vspace{-0.5cm}

\begin{thebibliography}{99}

\bibitem{Greiner02} M. Greiner, O. Mandel, T. Esslinger,
  T.W. H\"ansch, and I. Bloch, Nature {\bf 415}, 39 (2002).

\bibitem{Stoferle04} T. St\"oferle, H. Moritz, C. Schori, M. K\"ohl,
  and T. Esslinger, Phys. Rev. Lett. {\bf 92}, 130401 (2004).

\bibitem{Fisher89}  M.P.A. Fisher, P.B. Weichman, G. Grinstein, and
  D.S. Fisher, Phys. Rev. B {\bf 40}, 546 (1989); D. Jaksch,
  C. Bruder, J.I. Cirac, C.W. Gardiner, and P. Zoller,
  Phys. Rev. Lett. {\bf 81}, 3108 (1998).

\bibitem{Pitaevskii03} See, for instance, L. Pitaevskii and S. Stringari,
  {\it Bose-Einstein condensation} (Oxford University Press, 2003). 

\bibitem{Sheshadri93} K. Sheshadri, H.R. Krishnamurthy, R. Pandit, and
  T.V. Ramakrishnan, Europhys. Lett. {\bf 22}, 257 (1993).

\bibitem{VanOosten01} D. Van Oosten, P. van der Straten, and
  H.T.C. Stoof, Phys. Rev. A {\bf 63},053601 (2001). 

\bibitem{Sachdev99} S. Sachdev, {\it Quantum Phase
  Transitions} (Cambridge University, Cambridge, England, 1999). 

\bibitem{Freericks94} J.K. Freericks and H. Monien, Europhys. Lett.
{\bf 26}, 545 (1994); {\it ibid.} Phys. Rev. B {\bf 53}, 2691
(1996).

\bibitem{Pairault98} S. Pairault, D. S\'en\'echal, and A.-M. S.
Tremblay, Phys. Rev. Lett. {\bf 80}, 5389 (1998); {\it ibid.},
Eur. Phys. J. B {\bf 16}, 85 (2000). 

\bibitem{Dupuis00} N. Dupuis, Nucl. Phys. B {\bf 617}, 618 (2000);
{\it ibid} cond-mat/0105063.

\bibitem{Negele} see, for instance, J.W. Negele and H. Orland, {\it
Quantum Many Particle Systems} (Addison-Wesley, 1988). 

\bibitem{note2} Here we neglect the ``anomalous'' terms of the action
  since they play no role in the following. \cite{Dupuis00}

\bibitem{note1} As noted by Pairault {\it et. al.}, \cite{Pairault98}
strong-coupling expansions usually lead to a non-physical
single-particle Green's function (e.g. a negative spectral weight in
fermion systems or a non-normalized spectral function). 

\bibitem{Konabe04} S. Konabe, T. Nikuni, and M. Nakamura, cond-mat/0407229. 

\bibitem{Dickerscheid03} D.B.M. Dickerscheid, D. van Oosten, P.J.H. Denteneer,
  and H.J. Stoof, Phys. Rev. A {\bf 68}, 043623 (2003). 

\end{thebibliography}
\end{document}